\RequirePackage{amsmath}
\documentclass[runningheads,a4paper]{llncs}

\input glyphtounicode.tex %

\usepackage[utf8]{inputenc}
\usepackage{amsfonts}
\usepackage{graphicx}
\usepackage{subfigure}
\usepackage{float}

\usepackage[retainorgcmds]{IEEEtrantools}
\usepackage[usenames,dvipsnames]{xcolor}
\usepackage{color}

\usepackage{bm}
\usepackage{hyperref}

\renewcommand{\figurename}[1]{Figure\,\,#1}
\renewcommand{\tablename}[1]{Table #1}

\begin{document}

\mainmatter  

\title{High Throughput Computation of Reference Ranges of Biventricular Cardiac Function on \\ the UK Biobank Population Cohort}

\titlerunning{Automatic Reference Ranges of Cardiac Function on the UK Biobank}

\authorrunning{R.~Attar et al.}
\author{Rahman~Attar$^1$, Marco~Perea\~nez$^1$, Ali Gooya$^1$, X{\`e}nia~Alb{\`a}$^2$, Le~Zhang$^1$, Stefan~K.~Piechnik$^3$, Stefan~Neubauer$^3$, Steffen~E.~Petersen$^4$, Alejandro~F.~Frangi$^1$}
\institute{$^1$Center for Computational Imaging and Simulation Technologies in Biomedicine, Department of Electronic and Electrical Engineering, The University of Sheffield, UK.\\$^2$Center for Computational Imaging and Simulation Technologies in Biomedicine, Universitat Pompeu Fabra, Barcelona, Spain.\\ $^3$Oxford Center for Clinical Magnetic Resonance Research (OCMR), Division of Cardiovascular Medicine, University of Oxford, John Radcliffe Hospital, UK.\\ $^4$Cardiovascular Medicine at the William Harvey Research Institute, Queen Mary University of London and Barts Heart Center, Barts Health NHS Trust, UK.}

\maketitle

\begin{abstract}
The exploitation of large-scale population data has the potential to improve healthcare by discovering and understanding patterns and trends within this data. To enable high throughput analysis of cardiac imaging data automatically, a pipeline should comprise quality monitoring of the input images, segmentation of the cardiac structures, assessment of the segmentation quality, and parsing of cardiac functional indexes. We present a fully automatic, high throughput image parsing workflow for the analysis of cardiac MR images, and test its performance on the UK Biobank (UKB) cardiac dataset. The proposed pipeline is capable of performing end-to-end image processing including: data organisation, image quality assessment, shape model initialisation, segmentation, segmentation quality assessment, and functional parameter computation; all without any user interaction. To the best of our knowledge, this is the first paper tackling the fully automatic 3D analysis of the UKB population study, providing reference ranges for all key cardiovascular functional indexes, from both left and right ventricles of the heart. We tested our workflow on a reference cohort of 800 healthy subjects for which manual delineations, and reference functional indexes exist. Our results show statistically significant agreement between the manually obtained reference indexes, and those automatically computed using our framework.
\end{abstract}


\section{Introduction}
Cardiovascular diseases (CVDs) are recognised as the number one cause of death worldwide \cite{roth2017global}. Diagnosis of cardiovascular disease is often made at late symptomatic stages, which leads to late interventions and decreased efficacy of medical care. Thus, mechanisms for early and reliable quantification of cardiac function is of utmost importance.

Analysis and interpretation of cardiac structural and functional indexes in large-scale population image data can reveal patterns and trends across population groups, and allow insights into risk factors before CVDs develop. UKB is one of the world’s largest population-based prospective studies, established to investigate the determinants of disease. 

In terms of population sample size, experimental setup, and quality control, the most reliable reference ranges for cardiovascular structure and function in adult caucasians aged 45-74 found in the literature are those reported in \cite{petersen2017reference}. In \cite{petersen2017reference}, cardiovascular magnetic resonance (CMR) scans were manually delineated and analysed using cvi42 post-processing software (Version 5.1.1, Circle Cardiovascular Imaging Inc., Calgary, Canada). These reference values are used in this paper to validate the proposed workflow.

In this paper, we present a fully automatic 3D image parsing workflow with quality control modules to analyse CMR images in the UKB and corroborate their validity compared to their manual counterpart. The proposed workflow is capable of segmenting the cardiac ventricles and generating clinical reference ranges that are statistically comparable to those obtained by human observers. The main contribution of this paper is in its clinical impact, resulting from the analysis of left ventricle (LV) and right ventricle (RV) of the heart, as well as the extraction of key cardiac functional indexes from large CMR datasets.
 
\section{Methods}
\begin{figure}[h]
	\centering\includegraphics[width=1\linewidth]{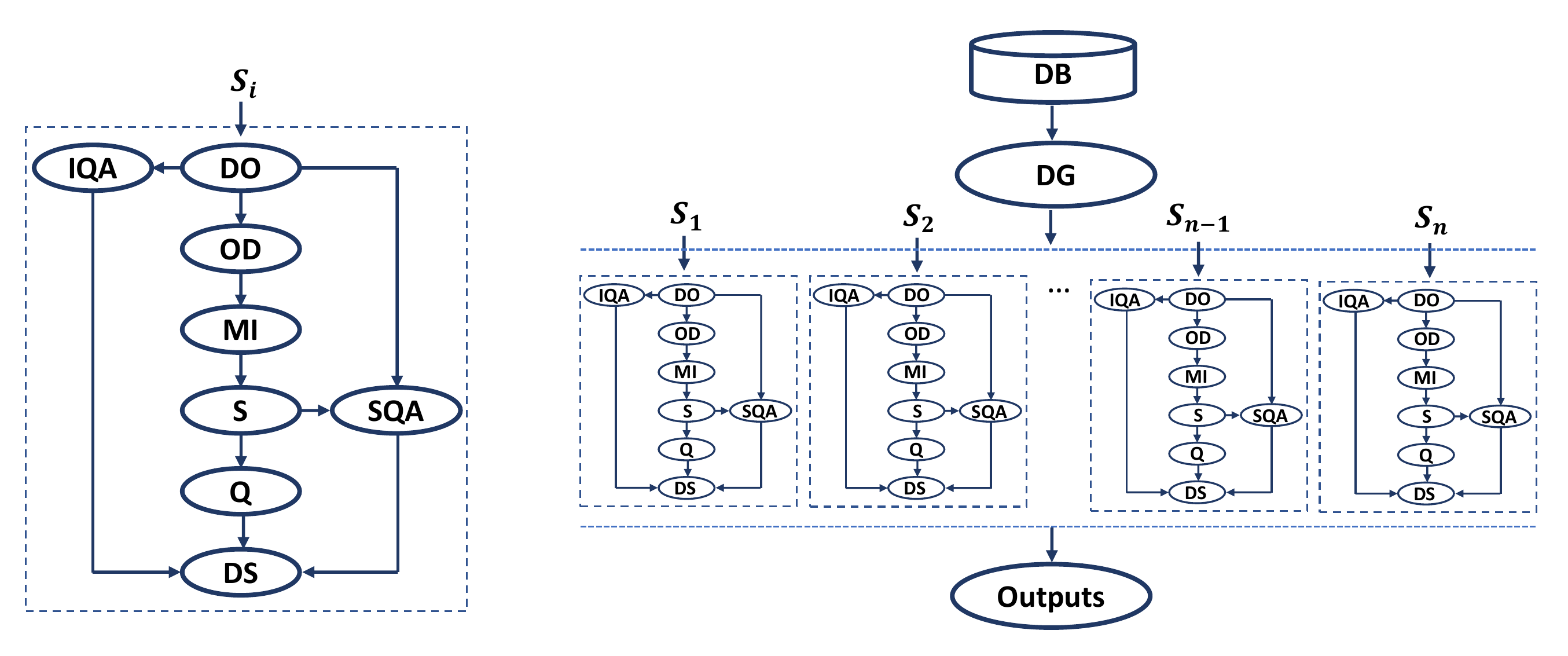}
	\caption{The proposed fully automatic image parsing workflow for the analysis of cardiac ventricles in parallel. Left: The workflow includes the following modules: DO: Data Organisation, IQA: Image Quality Assessment, OD: Organ Detection, MI: Model Initialisation, S: Segmentation, SQA: Segmentation Quality Assessment, Q: Quantification, DS: Data Sink. Right: The quantitative functional analysis of a large database in parallel mode. DB: Database, DG: Data Grabber, n: number of subjects, and $S_{i}$: $i^{th}$ subject of the dataset.}
	\label{fig:pipeline}
\end{figure}

Figure \ref{fig:pipeline} shows the architecture of the proposed workflow addressing the issue of large-scale analysis of CMR images. It consists of eight main modules to analyse every single subject of the database. To create a modular workflow and enable processing multiple subjects in parallel, a workflow manager software package is required. This provides an infrastructure for the set-up, performance and monitoring of a defined sequence of tasks, regardless their programming language. In our implementation, the Nipype package \cite{gorgolewski2011nipype} has been used. It allows us to combine a heterogeneous set of software packages within a single and highly efficient workflow, processing several subjects in parallel using cloud computing platforms provided by Amazon (high performance processors and S3 storage services).

\subsection{Data Organisation (DO)}
The Data Organisation (DO) module was developed to hierarchically organise image series from raw DICOM data. It is important to organise the data files to minimise redundancy and inconsistency. As a result, the organised data provides improved searchability and identification of contents. Clear, descriptive, and unique file names have been used to reflect the contents of the file, uniquely identify the data, and enable precise accessibility and data retrieval. Each subject's DICOM data are organised according to cardiac cycle phase, and into short axis (SAX) and long axis (LAX) views.

\subsection{Image Quality Assessment (IQA)}
Low image quality can not be fully avoided, particularly in large-scale imaging studies. To ensure that the quality of collected data is optimal for statistical analysis, having an IQA module is of paramount. This allows the automatic detection of abnormal images, whose analysis would otherwise impair the aggregated statistics over the cohort. Since the lack of basal and/or apical slices is the most common problem affecting image quality in CMR images, and has a major impact on the accuracy of quantitative parameters of cardiac function, our IQA module is designed to detect missing apical and basal slices of the CMR input. Thus, every top and bottom short axis view of input image volumes is analysed using two convolutional neural networks, each particularly trained for detection of missing slices in the basal and/or apical positions. The details of the architecture used can be found in \cite{zhang2016automated}.

\subsection{Organ Detection (OD)}
\label{OD}
To segment the image, we use a Sparse Active Shape Models framework (SPASM) \cite{van2006spasm}, which requires model initialisation. We achieve this automatically by extending the method proposed in \cite{alba2017automatic} for LV initialisation to biventricular initialisation. In \cite{alba2017automatic}, the location of the LV is determined by a rough estimation of the intersection of slices from different views (SAX and LAX).  Then, a Random Forest regressor trained with two complementary feature descriptors (i.e. the Histogram of Oriented Gradients and Gabor Filters) is used to predict the final landmark positions. This method is LV specific and therefore we have extended it to take into account image features corresponding to the RV, and obtain optimal initialisations for biventricular segmentation.

\subsection{Model Initialization (MI)}
The landmarks obtained in Sec. \ref{OD} are used 1) to suitably place the initial shape inside the image volume (translation), 2) to scale the initial shape along the main axis of the heart (scaling); and 3) to define the initial orientation of the heart based on the relative position of the mitral valve (rotation). These initial pose parameters are estimated by registering the obtained landmarks to their corresponding points on the mean model shape. As we segment all timepoints in the CINE sequence, we initialise the first image timepoint with the model mean, however, subsequent cardiac phases are initialised with the resulting segmentation from the previous timepoint.

\subsection{Segmentation (S)}
Cardiac LV and RV segmentation is performed with a modified 3D-SPASM segmentation method \cite{van2006spasm}. The main components of the 3D-SPASM are a Point Distribution Model (PDM), an Intensity Appearance Model (IAM), and a Model Matching Algorithm (MMA). 

In this work, the PDM is a surface mesh representing the endocardial and epicardial surfaces for the LV and the endocardial surface for the RV. The PDM is built during training by applying Principal Component Analysis to a set of aligned shapes and maintaining eigenvectors corresponding to a predefined percentage of shape variability. The learned shape variability can be modeled as ${\mathbf{\hat x}} = {\mathbf{\bar x}} + \mathbf{\Phi b}$ where ${\mathbf{\hat x}}$ is a shape model instance, $\mathbf{\bar{x}}$ is the mean shape, $\mathbf{\Phi}$ is an eigenvector matrix and $\mathbf{b}$ is a vector of scaling values for each principal component. By modifying $\mathbf{b}$, we can generate shapes from the shape distribution.

The IAM is trained by learning the graylevel intensity distribution along perpendiculars to boundary points on the cardiac shape. An appearance mean and covariance matrix $\{ \mathbf{  \bar g }, \mathbf{  \Sigma  }_{gg} \}$ is computed for each landmark by sampling the intensity around each point over the image training set. 

The last element of the segmentation process is the MMA, whose role is iterating between finding the optimal location of boundary points by distance minimisation between sampled image profiles and the IAM, and projection of these points onto the valid shape space defined by the PDM.

\subsection{Segmentation Quality Assessment (SQA)}
Due to varying image quality, image artefacts, or extreme anatomical variations found in large-scale studies it is essential to have a self-verification capabilities to automatically detect incorrect results, either to reprocess those images, or disregard them. This becomes even more important when automated segmentation methods are applied to large-scale datasets, and the segmentation results are to be used for further statistical population analysis \cite{valindria2017reverse}.
In our pipeline we incorporate the SQA proposed in \cite{alba2017automatic}. The SQA uses Random Forest classifiers trained on intensity features associated to blood pool and myocardium, and is able to detect successful segmentations.

\subsection{Quantification (Q)}
After successful SQA, we compute a thorough set of functional parameters based on blood-pool and myocardial volumes. To reproduce the reference ranges reported in \cite{petersen2017reference}, our quantification module performs volume computations using the Simpson’s rule. The principle underlying this method is that total volume can be approximated by the summation of stacks of elliptical disks.

\section{Experiments and Results}
We use the same dataset exploited in \cite{petersen2017reference}, and evaluate the performance of the proposed automatic workflow in two ways: 1) applying common metrics for segmentation accuracy assessment i.e., Dice Similarity Coefficient (DSC), Mean Contour Distance (MCD) and Hausdorff distance (HD), against ground truth values obtained through manual delineation by clinicians. 2) comparing cardiac biventricular function indexes derived from manual and automatic segmentations such as ventricular end diastolic/systolic volumes and myocardial mass. Additionally, quantitative evaluation of human performance i.e., the inter-observer variability, is measured among the manual segmentations of different clinical experts. A set of 50 subjects was randomly selected and each subject was analysed by three expert observers (O1, O2, O3) independently. We compare the result of segmentation on the same set of subjects to show how close the performance of the automatic segmentation is to human performance and also the performance of the proposed workflow on a large dataset.

Image volumes at end diastolic and end systolic timepoints of 250 random subjects (500 images in total) were used for training the PDM and IAM. The test dataset contains 800 subjects (not included in training) used for evaluation of the proposed automatic approach. The input images and output segmentation contours were automatically quality controlled to ensure that image volumes included both basal and apical slices, and to verify the automatic segmentation results. After IQA, all 800 images were classified as having full coverage. After SQA, 21 segmentations were deemed suboptimal. Since the aim of the results presented in Sec. \ref{accu} is the evaluation of segmentation accuracy, all 800 segmentation results (including 21 outliers) were included in the statistics. In contrast, those results presented in Sec. \ref{CFI} are based on 779 good quality segmentations, i.e. excluding those deemed suboptimal by SQA.

\subsection{Segmentation Accuracy} \label{accu}
Table \ref{DSCMCDHD} reports mean and standard deviation for DSC, MCD, and HD comparing between automatic and manual segmentations performed on test sets of 50 and 800 subjects never seen before by the PDM and IAM. The set of 50 subjects is the same set used for the evaluation of inter-observer variability. The set of 800 subjects is the same set used to generate reference ranges in \cite{petersen2017reference}. 

The reported DSC values show excellent agreement ($\geq 0.87$) between manual delineations and automatic segmentations. MCD errors are smaller than the in-plane pixel spacing range of 1.8 mm to 2.3 mm found in the UKB. Although HD is larger than the in-plane pixel spacing, it is still within an acceptable range when compared with the distance range seen  between different human observers.
Table \ref{DSCMCDHD} shows that the segmentation accuracy of our method is within error ranges observed between different human raters. This indicates that our workflow performs with human-like reliability, and can fully automatically segment large scale datasets where manual inputs are infeasible.
\vspace{-0.4cm}

\begin{table*}[ht]

	\caption{Segmentation accuracy expressed in terms of DSC, MCD and HD comparing the automatic (Auto.), manual (Man.), and observers (O1-O3) segmentations. \newline LV\textsubscript{endo}: LV endocardium. LV\textsubscript{myo}: LV myocardium, RV\textsubscript{endo}: RV endocardium. Values indicate mean $\pm$ standard deviation.}
	\centering
	
	\smallskip
	\scalebox{0.9}{
		\begin{tabular}{l c c c | c c }
			\multicolumn{6}{@{}l}{(a) DSC}                                                                                           \\ \hline
			                      & O1 vs O2        & O2   vs   O3        & O3   vs   O1        & Auto. vs  Man.        & Auto. vs Man.                  \\
			                      & (n=50)          & (n=50)          & (n=50)          & (n=50)         & (n=800)                   \\ \hline
			LV\textsubscript{endo} & 0.94 $\pm$ 0.04 & 0.92 $\pm$ 0.04 & 0.93 $\pm$ 0.04 &  \textbf{0.93 $\pm$ 0.03} & \textbf{0.93 $\pm$ 0.04} \\
			LV\textsubscript{myo} & 0.88 $\pm$ 0.02 & 0.87 $\pm$ 0.03 & 0.88 $\pm$ 0.02 &  \textbf{0.88 $\pm$ 0.03} & \textbf{0.87 $\pm$ 0.03} \\
			RV\textsubscript{endo} & 0.87 $\pm$ 0.06 & 0.88 $\pm$ 0.05 & 0.89 $\pm$ 0.05 &  \textbf{0.87 $\pm$ 0.06} & \textbf{0.89 $\pm$ 0.05} \\ \hline
		\end{tabular}}
		
		\smallskip  
		\scalebox{0.9}{
			\begin{tabular}{l c c c | c c }
				\multicolumn{6}{@{}l}{(b) MCD (mm)}                                                                                      \\ \hline
				                      & O1 vs O2        & O2 vs O3        & O3 vs O1        & Auto. vs  Man.        & Auto. vs  Man.                 \\
				                      & (n=50)          & (n=50)          & (n=50)          & (n=50)         & (n=800)                   \\ \hline
				LV\textsubscript{endo} & 1.00 $\pm$ 0.25 & 1.30 $\pm$ 0.37 & 1.21 $\pm$ 0.48 & \textbf{1.28 $\pm$ 0.39} & \textbf{1.17 $\pm$ 0.32} \\
				LV\textsubscript{myo} & 1.16 $\pm$ 0.34 & 1.19 $\pm$ 0.25 & 1.21 $\pm$ 0.36 & \textbf{1.20 $\pm$ 0.34} & \textbf{1.16 $\pm$ 0.40} \\
				RV\textsubscript{endo} & 2.00 $\pm$ 0.79 & 1.78 $\pm$ 0.45 & 1.87 $\pm$ 0.74 & \textbf{1.79 $\pm$ 0.80} & \textbf{1.81 $\pm$ 0.67} \\ \hline
			\end{tabular}}

			\smallskip  
			\scalebox{0.9}{
				\begin{tabular}{l c c c | c c }
					\multicolumn{6}{@{}l}{(c) HD (mm)}                                                                                       \\ \hline
					                      & O1 vs O2        & O2 vs O3        & O3 vs O1        & Auto. vs  Man.        & Auto. vs Man.                  \\
					                      & (n=50)          & (n=50)          & (n=50)          & (n=50)         & (n=800)                   \\ \hline
					LV\textsubscript{endo} & 2.84 $\pm$ 0.70 & 3.31 $\pm$ 0.90 & 3.25 $\pm$ 0.96 & \textbf{3.21 $\pm$ 0.97} & \textbf{3.21 $\pm$ 0.99} \\
					LV\textsubscript{myo} & 3.70 $\pm$ 1.16 & 3.82 $\pm$ 1.07 & 3.76 $\pm$ 1.21 & \textbf{3.91 $\pm$ 1.20} & \textbf{3.92 $\pm$ 1.30} \\
					RV\textsubscript{endo} & 7.56 $\pm$ 5.51 & 7.35 $\pm$ 2.19 & 7.14 $\pm$ 2.20 & \textbf{7.41 $\pm$ 4.11} & \textbf{7.31 $\pm$ 3.32} \\ \hline
				\end{tabular}}
					\label{DSCMCDHD}
\end{table*}
\vspace{-0.8cm}

\subsection{Estimation of Cardiac Function Indexes} \label{CFI}
We evaluate the accuracy of cardiac function indexes derived from automatic segmentation versus gold standard reference ranges derived from manual segmentation. We calculate the LV end-diastolic volume (LVEDV) and end-systolic volume (LVESV), LV Stroke Volume (LVSV), LV Ejection-Fraction (LVEF), LV myocardial mass (LVM), RV end-diastolic volume (RVEDV) and end-systolic volume (RVESV), RV Stroke Volume (RVSV) and RV Ejection-Fraction (RVEF) from automated segmentation and compare them to measurements from manual segmentation. 

Table \ref{reference} shows excellent agreement between the mean and standard deviation of ventricular parameters of a healthy population obtained through both automatic and manual segmentations. Furthermore, we performed two-sample Kolmogorov-Smirnov (K-S) tests to show that ventricular parameters obtained through manual and automatic approaches are drawn from the same population, under the null hypothesis that the manual and automatic methods are from the same continuous distribution in terms of clinical indexes. K-S test on different indexes does not reject the null hypothesis of being from same distribution at the 5\% significance level.
\vspace{-0.4cm}

\begin{table*}[!ht]
	\centering
	\caption{Cardiac function indexes derived from manual (Man.) vs automatic (Auto.) segmentation on 779 subjects. Values indicate mean $\pm$ standard deviation.}
	\label{my-label}
				\scalebox{1}{
	\begin{tabular}{ll|l|l|l|l|l|l|l|l} \hline
		\multicolumn{1}{l}{} & \multicolumn{1}{c|}{\begin{tabular}[c]{@{}c@{}}
			LVEDV \\
			(ml)
		\end{tabular}} & \multicolumn{1}{c|}{\begin{tabular}[c]{@{}c@{}}LVESV\\ (ml)\end{tabular}} & \multicolumn{1}{c|}{\begin{tabular}[c]{@{}c@{}}LVSV\\ (ml)\end{tabular}} & \multicolumn{1}{c|}{\begin{tabular}[c]{@{}c@{}}LVEF\\ (\%)\end{tabular}} & \multicolumn{1}{c|}{\begin{tabular}[c]{@{}c@{}}LVM\\ (g)\end{tabular}} & \multicolumn{1}{c|}{\begin{tabular}[c]{@{}c@{}}RVEDV\\ (ml)\end{tabular}} & \multicolumn{1}{c|}{\begin{tabular}[c]{@{}c@{}}RVESV\\ (ml)\end{tabular}} & \multicolumn{1}{c|}{\begin{tabular}[c]{@{}c@{}}RVSV\\ (ml)\end{tabular}} & \multicolumn{1}{c}{\begin{tabular}[c]{@{}c@{}}RVEF\\ (\%)\end{tabular}} \\ \hline
	
		Man. & 144 $\pm$ 34  &  59 $\pm$ 18  &  85 $\pm$ 20  & 60 $\pm$ 6  & 86 $\pm$ 24 & 154 $\pm$ 40 & 69 $\pm$ 24  & 85 $\pm$ 20 & 56 $\pm$ 6     \\ 
		Auto. & 146 $\pm$ 31  &  60 $\pm$ 18  &  86 $\pm$ 18  & 60 $\pm$ 7  & 87 $\pm$ 23 & 154 $\pm$ 40 & 71 $\pm$ 26  & 83 $\pm$ 21 & 54 $\pm$ 7 \\ \hline
	\end{tabular}}
	\label{reference}
\end{table*}

\begin{figure*}[ht]

	\centering\includegraphics[width=1.04\textwidth]{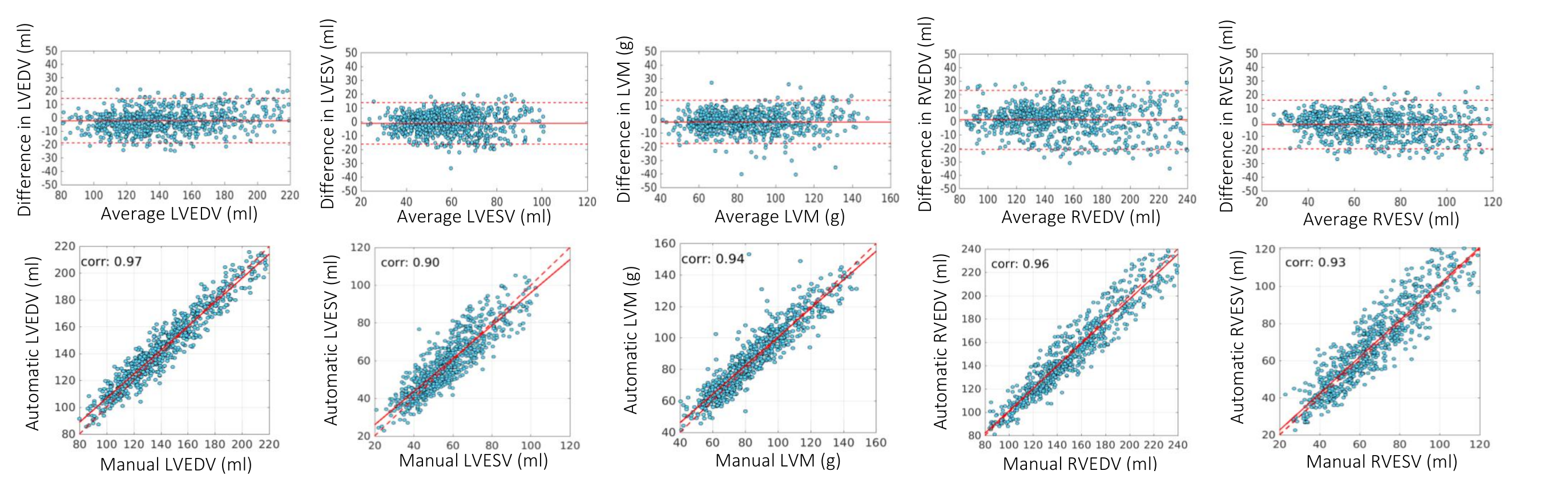}
	\caption{Repeatability of various cardiac functional indexes: manual vs automatic analysis on the test dataset. The first row shows \textbf{Bland-Altman} plots. The solid line denotes the mean difference (bias) and the two dashed lines denote $\pm$1.96 standard deviations from the mean. The second row shows \textbf{correlation} plots. The dashed and solid line denote the identity and linear regression lines, respectively.}
		\label{BAcorr}
\end{figure*}

Figure \ref{BAcorr} shows Bland-Altman (top) and correlation (bottom) plots of ventricular parameters computed using the proposed automatic method and the manual reference on the test dataset. The Bland-Altman plots show good limits of agreement and also the mean difference line nearly at zero, which suggests that the clinical indexes obtained through the automatic approach have little bias. The correlation plots and their correlation coefficient (corr) indicate a strong relationship between the manual and automatic approaches. 

\section{Conclusion}
In this paper, we propose a fully automatic workflow capable of performing high throughput end-to-end 3D cardiac image analysis. We tested our workflow on a reference cohort of 800 healthy subjects for which manual delineations, and reference functional indexes exist. Our results show statistically significant agreement between the manually obtained reference indexes, and those computed automatically using the proposed workflow. As future work, we plan to analyse all available UKB datasets including both healthy and pathological subjects and report the regional and global cardiac function indexes.

\noindent
\newline
\textbf{{\large Acknowledgements}} \hspace{0.1cm} R. Attar was funded by the Faculty of Engineering Doctoral Academy Scholarship, University of Sheffield. This work has been partially supported by the MedIAN Network (EP/N026993/1) funded by the Engineering and Physical Sciences Research Council (EPSRC), and the European Commission through FP7 contract VPH-DARE@IT (FP7-ICT-2011-9-601055) and H2020 Program contract InSilc (H2020-SC1-2017-CNECT-2- 777119). The UKB CMR dataset has been provided under UK Biobank Application 2964.

\bibliographystyle{ieeetr}
\bibliography{references}

\end{document}